\documentclass[12pt]{article}
\usepackage{amssymb}

\usepackage{amsmath}
\usepackage{cmmib57}

\def\be{\begin{equation}}
\def\ee{\end{equation}}
\def\ba{\begin{eqnarray}}
\def\ea{\end{eqnarray}}

\begin{document}

\title{\textbf{Brane World Cosmology In Jordan-Brans-Dicke Theory }}
\author{\textbf{M. Ar\i k}\thanks{
email: arikm@boun.edu.tr}, \textbf{D. \c{C}iftci}\thanks{%
email: dciftci@boun.edu.tr}, \\
%EndAName
Bo\u{g}azi\c{c}i University, Department of Physics,\\
34342 Bebek/ Istanbul TURKEY. }
\date{\today}
\maketitle

\begin{abstract}
We consider the embedding of 3+1 dimensional cosmology in 4+1 dimensional
Jordan-Brans-Dicke theory. We show that exponentially growing and power law
scale factors are implied. Whereas the 4+1 dimensional scalar field is
approximately constant for each, the effective 3+1 dimensional scalar field
is constant for exponentially growing scale factor and time dependent for
power law scale factor.\vspace{0.2in}
\end{abstract}

%\begin{quotation}
%{BRANE WORLD COSMOLOGY IN JORDAN-BRANS-DICKE THEORY} \vspace{0.11in}

%\bigskip

%{\huge \vspace{0.05in}\allowbreak } {M. ARIK* and D. \c{C}\.{I}FTC\.{I}**}

%\allowbreak \medskip {\small Bo\u{g}azi\c{c}i University, Department of
%Physics, Istanbul, Turkey}

%\bigskip

%*arikm@boun.edu.tr

%**dciftci@boun.edu.tr

%\ \hspace{0.3in}
%{\small \textbf{Abstract}.

%\end{quotation}

\section*{Introduction}

%\begin{center} {INTRODUCTION} \vspace{0.11in}
%\end{center}

In spite of the success of general relativity now called the standard theory
of gravitation, there are many other alternative theories. Among them the
scalar tensor theory is the most important one. The scalar-tensor theory was
conceived originally by P. Jordan \cite{Jordan} who started to embed a four
dimensional curved manifold in five dimensional flat space-time. He
presented a general Lagrangian for the scalar field living in
four-dimensional curved space-time:%
\begin{equation}
L_{J}=\sqrt{-g}\left[ \varphi _{J}^{\gamma }\left( R-\omega _{J}\frac{1}{%
\varphi _{J}^{2}}g^{\mu \nu }\partial _{\mu }\varphi _{J}\partial _{\nu
}\varphi _{J}\right) +L_{matter}\left( \varphi _{J},\Psi \right) \right] ,
\end{equation}%
where $\varphi _{J}\left( x\right) $ is Jordan's scalar field, $\gamma $ and
$\omega _{J}$ are constants, and $\Psi $ represents matter fields. $\varphi
_{J}^{\gamma }R$ is the nonminimal coupling term which marked the birth of
scalar-tensor theory.

\qquad Jordan's work was taken over particularly by C.Brans and R.H. Dicke %
\cite{Brans}. They assumed that decoupling of scalar field from the matter
part of the Lagrangian occurs. They defined their scalar field $\varphi $ by%
\begin{equation}
\varphi =\varphi _{J}^{\gamma }
\end{equation}%
and then Lagrangian will be
\begin{equation}
L_{BD}=\sqrt{-g}\left( \varphi R-\omega \frac{1}{\varphi }g^{\mu \nu
}\partial _{\mu }\varphi \partial _{\nu }\varphi +L_{matter}\left( \Psi
\right) \right) .
\end{equation}%
They demanded that the matter part of the Lagrangian $\sqrt{-g}L_{matter}$
be decoupled from $\varphi \left( x\right) $ as their requirement that the
weak equivalence principle be respected, in contrast to Jordan's model \cite%
{Yasunory}. To remove the singularity from the second term on the right hand
side we introduce a new field $\phi $:%
\begin{equation}
\varphi =\frac{\phi ^{2}}{8\omega }.
\end{equation}%
Then the Brans-Dicke (B-D) action will be%
\begin{equation}
L_{BD}=\sqrt{-g}\left( \frac{\phi ^{2}}{8\omega }R-\frac{1}{2}g^{\mu \nu
}\partial _{\mu }\phi \partial _{\nu }\phi +L_{matter}\right) .
\end{equation}%
where in order to get cosmic acceleration, either the parameter
$\omega $ should be time dependent \cite{Banerjee}, or a potential
term for the scalar field could be added \cite{Sen} to the
Lagrangian.

On the other hand, string theory predicts a new type of nonlinear structure,
which is called a brane, a word created from ''membrane''. This also gives a
new perspective to cosmology so that our universe is confined to a four
dimensional space-time subspace or 3-brane. The extra dimension may have
large compact toroidal topology \cite{Arkani} or be unbounded with a warp
factor, depending on the distance from the brane \cite{RS1,RS2}.
Additionally several works have studied higher dimensional B-D theory to
combine the advantages of both the five dimensional cosmology and the B-D
theory \cite{Freund,Mendes}. Moreover, considering the scalar field in the
five dimensional bulk with Einstein gravity was proposed by many works \cite%
{potantial}.

Our starting point is the paper of Bander who studied five dimensional bulk
whose dynamics is governed by a scalar Liouville field coupled to gravity in
the usual way \cite{Bander}. Then he derived that the effective theory on
the brane has a time dependent Planck mass and cosmological constant and
also found expanding scale factors with no acceleration. In this paper we
investigate the properties of the five dimensional bulk in Brans-Dicke
theory. The layout of our paper is as follows. In section 2 we present the
general framework for our five dimensional theory and compute the five
dimensional Brans-Dicke equations. In section 3 we analyze the cosmological
solutions. We find false vacuum energy ($p_{B}=-\rho _{B}$) for
exponentially growing scale factors and radiation dominated universe ($p_{B}=%
\frac{1}{3}\rho _{B}$) for power law scale factors in the bulk. In section 4
we derive the effective four dimensional scalar field and obtain its time
dependence. Finally, we sum up our results and conclusions in section 5. \ \

\section{ The Action and Equations of Motion}

In this work we look at the five dimensional Brans-Dicke action:
\begin{equation}
S=\int d^{5}x\sqrt{g}\left( \frac{\phi ^{2}}{8\omega }R-\frac{1}{2}\partial
_{A}\phi \partial _{B}\phi g^{AB}-V\left( \phi \right) \right) ,
\label{action}
\end{equation}%
where $\omega $ is the dimensionless Brans-Dicke parameter, $\phi $ is the
scalar field and $V\left( \phi \right) $ is the scalar potential. The
variation of the action with respect to $g^{AB}$ gives%
\begin{equation}
\frac{1}{8\omega }\left( \phi ^{2}G_{AB}-\phi _{,A;B}^{2}+g_{AB}\square \phi
^{2}\right) -\frac{1}{2}\partial _{A}\phi \partial _{B}\phi +\frac{g_{AB}}{4}%
\partial _{C}\phi \partial ^{C}\phi +\frac{1}{2}g_{AB}V\left( \phi \right)
=T_{AB}.  \label{gmn}
\end{equation}%
We choose a general five dimensional metric anzats which can be written in
an orthonormal basis as \cite{Bander}:%
\begin{equation}
ds^{2}=b\left( t\right) ^{2}dW^{2}+f\left( W\right) ^{2}\left[
-dt^{2}+a\left( t\right) ^{2}\delta _{ij}dx^{i}dx^{j}\right] ,  \label{dsb-d}
\end{equation}%
where $i,j=1,2,3$ and $f\left( W\right) $ is the warp factor which depends
on the fifth coordinate, $a\left( t\right) $ is the cosmological scale
factor and $b\left( t\right) $ is the time dependent scale factor of the
fifth dimension. More generally, this metric has been studied in papers \cite%
{Mendes,Deffayet}. In Mendes' work \cite{Mendes} the five dimensional brane
cosmology with non-minimally coupled scalar field to gravity is interpreted
in Jordan frame without a scalar potential. In our work we add a scalar
potential to the action.

In the orthonormal basis $e^{0}=fdt,$ $e^{i}=fadx^{i}$ and $e^{5}=bdW,$ the
stress-energy tensor can be considered as \cite{Deffayet}
\begin{equation}
T_{B}^{A}=T_{B}^{A}\mid _{bulk}+T_{B}^{A}\mid _{brane},
\end{equation}%
where $T_{B}^{A}\mid _{bulk}$is the energy momentum tensor of the bulk
matter and%
\begin{equation*}
T_{B}^{A}\mid _{bulk}=\text{diag}\left( -\rho
_{B},p_{B},p_{B},p_{B},q_{B}\right) .
\end{equation*}%
The second term corresponds to the matter content in the brane $\left(
W=0\right) ,$%
\begin{equation*}
T_{B}^{A}\mid _{brane}=\frac{\delta \left( W\right) }{b}\text{diag}\left(
-\rho ,p,p,p,0\right) .
\end{equation*}%
If we substitute the Einstein tensor components in eq(\ref{gmn}) we obtain
the B-D equations. In the coordinate basis, for component $00;$

\begin{eqnarray}
&&\frac{1}{8\omega }\left( 3\left( \frac{\dot{a}^{2}}{a^{2}}+\frac{\dot{a}%
\dot{b}}{ab}\right) -\frac{3f^{2}}{b^{2}}\left( \frac{\acute{f}^{2}}{f^{2}}+%
\frac{\text{\H{f}}}{f}\right) +3\frac{\dot{a}}{a}\frac{\partial _{t}\phi ^{2}%
}{\phi ^{2}}+\frac{\dot{b}}{b}\frac{\partial _{t}\phi ^{2}}{\phi ^{2}}-\frac{%
f^{2}}{b^{2}}\left( 3\frac{\acute{f}}{f}\frac{\partial _{W}\phi ^{2}}{\phi
^{2}}+\frac{\partial _{W}^{2}\phi ^{2}}{\phi ^{2}}\right) \right)
\label{G00} \\
&&-\frac{f^{2}}{4b^{2}}\frac{\left( \partial _{W}\phi \right) ^{2}}{\phi ^{2}%
}-\frac{1}{4}\frac{\left( \partial _{t}\phi \right) ^{2}}{\phi ^{2}}-\frac{%
f^{2}}{2}\frac{V\left( \phi \right) }{\phi ^{2}}=\frac{T_{00}}{\phi ^{2}}.
\notag
\end{eqnarray}%
For components $ii;$%
\begin{eqnarray}
&&\frac{1}{8\omega }\left( -\left( \frac{2\ddot{a}}{a}+\frac{\dot{a}^{2}}{%
a^{2}}+2\frac{\dot{a}\dot{b}}{ab}+\frac{\ddot{b}}{b}\right) +\frac{3f^{2}}{%
b^{2}}\left( \frac{\acute{f}^{2}}{f^{2}}+\frac{\text{\H{f}}}{f}\right) -%
\frac{\partial _{t}^{2}\phi ^{2}}{\phi ^{2}}-\frac{\dot{b}}{b}\frac{\partial
_{t}\phi ^{2}}{\phi ^{2}}-2\frac{\dot{a}}{a}\frac{\partial _{t}\phi ^{2}}{%
\phi ^{2}}\right)  \label{Gii} \\
&&+\frac{1}{8\omega }\frac{f^{2}}{b^{2}}\left( \frac{\partial _{W}^{2}\phi
^{2}}{\phi ^{2}}+3\frac{\acute{f}}{f}\frac{\partial _{W}\phi ^{2}}{\phi ^{2}}%
\right) -\frac{1}{4}\frac{\left( \partial _{t}\phi \right) ^{2}}{\phi ^{2}}+%
\frac{f^{2}}{4b^{2}}\frac{\left( \partial _{W}\phi \right) ^{2}}{\phi ^{2}}+%
\frac{f^{2}}{2}\frac{V\left( \phi \right) }{\phi ^{2}}=\frac{1}{a^{2}}\frac{%
T_{ii}}{\phi ^{2}}.  \notag
\end{eqnarray}%
For component $55;$

\begin{eqnarray}
&&\frac{1}{8\omega }\left( -3\left( \frac{\dot{a}^{2}}{a^{2}}+\frac{\ddot{a}%
}{a}\right) +6\frac{f^{2}}{b^{2}}\frac{\acute{f}^{2}}{f^{2}}-\frac{\partial
_{t}^{2}\phi ^{2}}{\phi ^{2}}-3\frac{\dot{a}}{a}\frac{\partial _{t}\phi ^{2}%
}{\phi ^{2}}+\frac{4f^{2}}{b^{2}}\frac{\acute{f}}{f}\frac{\partial _{W}\phi
^{2}}{\phi ^{2}}\right)  \\
&&-\frac{f^{2}}{4b^{2}}\frac{\left( \partial _{W}\phi \right) ^{2}}{\phi ^{2}%
}-\frac{1}{4}\frac{\left( \partial _{t}\phi \right) ^{2}}{\phi ^{2}}+\frac{%
f^{2}}{2}\frac{V\left( \phi \right) }{\phi ^{2}}=\frac{f^{2}}{b^{2}}\frac{%
T_{55}}{\phi ^{2}}.  \notag  \label{55}
\end{eqnarray}%
For component $05;$%
\begin{equation}
\frac{1}{8\omega }\left( \frac{3\dot{b}\acute{f}}{bf}-\frac{\partial
_{t}\partial _{W}\phi ^{2}}{\phi ^{2}}+\frac{\acute{f}}{f}\frac{\partial
_{t}\phi ^{2}}{\phi ^{2}}+\frac{\dot{b}}{b}\frac{\partial _{W}\phi ^{2}}{%
\phi ^{2}}\right) -\frac{1}{2}\frac{\partial _{t}\phi \partial _{W}\phi }{%
\phi ^{2}}=0.  \label{05}
\end{equation}%
Assume that the $05$ component of the energy-momentum tensor vanishes, which
means that there is no flow of matter along the fifth dimension. Therefore
the nonzero elements of the 5D stress-energy tensor are%
\begin{eqnarray}
T_{00} &=&f^{2}\rho _{B}+f^{2}\frac{\delta \left( w\right) }{b}\rho  \\
T_{ii} &=&a^{2}f^{2}\rho _{B}+a^{2}f^{2}\frac{\delta \left( w\right) }{b}p
\notag \\
T_{55} &=&b^{2}q_{B}.  \notag
\end{eqnarray}%
Finally variation with respect to $\phi $ gives,%
\begin{equation*}
\frac{1}{4\omega }\left( \phi R\right) -\frac{\partial V\left( \phi \right)
}{\partial \phi }+\square \phi =0,
\end{equation*}%
which explictly reads%
\begin{equation}
\frac{1}{4\omega }R-\frac{\partial _{t}^{2}\phi }{f^{2}\phi }+\frac{4}{b^{2}}%
\frac{\acute{f}}{f}\frac{\partial _{W}\phi }{\phi }-\frac{3}{f^{2}}\frac{%
\dot{a}}{a}\frac{\partial _{t}\phi }{\phi }-\frac{\dot{b}}{bf^{2}}\frac{%
\partial _{t}\phi }{\phi }+\frac{\partial _{W}^{2}\phi }{b^{2}\phi }-\frac{1%
}{\phi }\frac{\partial V\left( \phi \right) }{\partial \phi }=0,  \label{f}
\end{equation}%
where the Ricci scalar $R$\ is:%
\begin{equation}
R=\frac{1}{f^{2}}\left( \frac{6\ddot{a}}{a}+\frac{2\ddot{b}}{b}+\frac{6\dot{a%
}^{2}}{a^{2}}+\frac{6\dot{a}\dot{b}}{ab}\right) -\frac{12\acute{f}^{2}}{%
f^{2}b^{2}}-\frac{8f^{\prime \prime }}{fb^{2}}.  \notag
\end{equation}%
The metric and the B-D field are continuous across the brane localized at $%
W=0$. However their derivatives can be discontinuous at the brane. Since we
have orbifold symmetry, second derivatives of scale factor and B-D field
will contain Dirac delta function in the second derivatives of the metric
with respect to fifth dimension. Therefore for a function $f,$ we have \cite%
{Mendes,Deffayet}
\begin{equation*}
f^{\prime \prime }=\widehat{f^{\prime \prime }}+\left[ f^{\prime }\right]
\delta \left( W\right) ,
\end{equation*}%
where $\widehat{f^{\prime \prime }}$ is the non-distributional part of the
double derivative of $f$, and $\left[ f^{\prime }\right] $ is the jump in
the first derivative of $f$ across $W=0$, it is defined as%
\begin{equation*}
\left[ f^{\prime }\right] =f^{\prime }\left( 0^{+}\right) -f^{\prime }\left(
0^{-}\right) .
\end{equation*}%
%
%
%
%
%.
Matching the Dirac delta functions in equations (\ref{G00}), (\ref{Gii}) and
(\ref{f}) we obtain that%
\begin{eqnarray}
\frac{\left[ f^{\prime }\right] _{0}}{f_{0}b_{0}} &=&-\frac{8\omega ^{2}}{%
\left( 3\omega +4\right) \phi ^{2}}\rho  \\
\frac{\left[ \phi ^{\prime }\right] _{0}}{\phi _{0}b_{0}} &=&-\frac{16\omega
}{\left( 3\omega +4\right) \phi ^{2}}\rho   \notag
\end{eqnarray}%
where the subscript `$0$' stands for the brane at $W=0.$ Using eq(\ref{G00})
and eq(\ref{Gii}) we get the remarkable result that the cosmological
constant dominates on the brane i.e.%
\begin{equation}
\rho =-p=-\frac{\phi _{0}^{2}}{8\omega }\left( \frac{3\left[ f^{\prime }%
\right] _{0}}{f_{0}b_{0}}+\frac{2\left[ \phi ^{\prime }\right] _{0}}{\phi
_{0}b_{0}}\right) .
\end{equation}%
Here choice of the scalar factor $a\left( t\right) $ does not make any
differences on the equation of state $\rho =-p.$ Using eq(\ref{05}) to
evaluate the jump condition we get the equation for the matter on the brane%
\begin{equation}
4\dot{\rho}+3\omega \frac{\dot{b}}{b}p+6\omega \frac{\dot{\phi}}{\phi }\rho
=0
\end{equation}%
where if $\ 2\frac{\dot{b}}{b}=\frac{\dot{\phi}}{\phi }$ or in particular
time derivatives of the $b$ and $\phi $ are zero, we obtain that $\rho $ and
$p=-\rho $ are constant on the brane.

\section{Solutions}

Solutions of B-D equations restrict the scalar field to be in the form $\phi
\left( t,W\right) =B\left( t\right) C\left( W\right) .$ Starting from this,
to satisfy all of the B-D equations we make two possible ansatze for $%
a\left( t\right) .$ The first one is exponential growth in time and the
other is power law expansion.

\subsection{\protect\bigskip Exponential Expansion, $a\left( t\right)
=a_{0}e^{\protect\lambda t}$}

We see from eq(\ref{G00}-\ref{f}) that $b\left( t\right) $ must be constant,
$b\left( t\right) =b_{0}$ and $B\left( t\right) $ must be in the exponential
form also $B\left( t\right) =B_{0}e^{\beta t}$ and than we can easily read
that%
\begin{equation}
f\left( W\right) =\frac{W}{W_{0}}.
\end{equation}%
For a brane at $W=W_{0}$, we introduce the coordinate $W^{\prime }$ such
that $\frac{W}{W_{0}}=1-\frac{W^{\prime }}{W_{0}}.$ The metric on both sides
of brane can be written
\begin{equation}
ds^{2}=b_{0}^{2}dW^{2}+\left( 1-\frac{\left| W\right| }{W_{0}}\right) ^{2}%
\left[ -dt^{2}+e^{2\lambda t}d\vec{x}^{2}\right] ,  \label{expds}
\end{equation}%
and the brane is at $W=0$. Here we dropped the prime for simplicity. This
warp factor is the same as in Bander's work \cite{Bander}. The brane we live
in is embedded in the five-dimensional bulk space-time and the four
dimensional part in the square parenthesis is the well known de-Sitter
space-time. This metric is similar to a Randall- Sundrum type of model in
the same sense. Instead of the exponential warp factor we obtain the linear
warp factor. However for small $W$ it is known that%
\begin{equation}
e^{-\left| W\right| }\simeq 1-\left| W\right| ,
\end{equation}%
and two the models are similar.

From equations (\ref{G00}) and (\ref{Gii}) it seems to be $p_{B}=-\rho _{B}$%
, which acts as a cosmological constant.\ In previous works \cite%
{Weinberg,Daila} this energy has been identified as the false vacuum energy
density $\rho _{f}.$ During the false vacuum phase the universe supercools.
It is believed that as the universe expands it cools down and then it
experiences a series of phase transitions. Since the cosmic expansion
continues to drive the temperature downward, the universe enters a period of
supercooling. As the universe supercools the energy density acts as an
effective cosmological constant. Therefore we can consider this stage as the
false phase.

For this condition we get the results
\begin{eqnarray}
C\left( W\right) &=&c_{0}\left( 1-\frac{\left| W\right| }{W_{0}}\right)
^{\alpha }, \\
\phi &=&B_{0}c_{0}\left[ e^{\beta t}\left( 1-\frac{\left| W\right| }{W_{0}}%
\right) \right] ^{\alpha },
\end{eqnarray}%
here $\phi $ depends only on the distance $W.$ Where on the brane we live
\begin{equation}
\left( 16\pi G\right) ^{-1}=M_{p}^{2}=\frac{\phi ^{2}}{8\omega }=\frac{%
\left( B_{0}c_{0}\right) ^{2}e^{2\alpha \beta }}{8\omega },  \label{g,mp,f}
\end{equation}%
where $B_{0}c_{0}$ is required to be within a few orders of magnitude of
Planck mass \cite{RS1}. We first discuss the solution where $T_{55}=0.$

\subsubsection{$T_{55}=0:$\emph{\ }}

From the B-D equations this condition causes $\rho _{B}=-p_{B}=0$ (empty
universe). Then solutions are very simple \newline

%[.cm]

\begin{center}
\begin{tabular}{|l|l|l|l|l|}
\hline
$\rho _{B}=0$ & $p_{B}=0$ & $V_{0}=0$ & $\beta =\lambda $ & $\alpha =\frac{1%
}{2\left( 1+\omega \right) }$ \\ \cline{3-5}
&  & $V_{0}=-\frac{\left( 3\omega +4\right) \lambda ^{2}}{2\omega \left(
1+\omega \right) ^{2}}\left( B_{0}c_{0}\right) ^{2/\alpha }$ & $\beta =0$ & $%
\alpha =\frac{1}{1+\omega }$ \\ \hline
\end{tabular}
\\[1cm]
\end{center}

\noindent where $b_{o}W_{o}\lambda =\pm 1.$ Here in the first row of the
table the B-D equations give a scalar field which depends not only on time
but also on the fifth coordinate. On the other hand in the second row the
scalar field only depends on the fifth coordinate and there is a scalar
potential $V_{0}\neq 0$. Therefore the scalar potential is not depend on the
time
\begin{equation*}
V\left( \phi \right) =V_{0}\phi ^{2-\frac{2}{\alpha }},
\end{equation*}%
where $V_{0}$ is a constant has dimension $L^{-2-\frac{3}{\alpha }}$. From
these results as $\omega \rightarrow \infty ,$ $\alpha ,V_{0}\rightarrow 0.$
Therefore $V\left( \phi \right) \rightarrow 0.$ This means that at the large
values of the B-D parameter, the scalar field is constant $\frac{\phi ^{2}}{%
8\omega }=M_{p}^{2}=\frac{\left( B_{0}c_{0}\right) ^{2}}{8\omega }$ with no
scalar potential.

The $q=0$ condition has been derived in \cite{biz} where it was found that
empty and flat five dimensional universe where $^{\left( 5\right) }R_{%
\phantom{MN}PQ}^{MN}=0$ and $\Lambda _{5}=0$ gives rise to a four
dimensional expanding universe with nonzero Riemann tensor and cosmological
constant. This five dimensional space is a well known Minkowski universe
\begin{equation}
ds^{2}=-dx_{1}^{2}+dx_{2}^{2}+dx_{3}^{2}+dx_{4}^{2}+dx_{5}^{2}
\end{equation}%
transformed into%
\begin{equation}
ds^{2}=b_{0}^{2}dW^{2}+\left( 1-\frac{\left| W\right| }{W_{0}}\right) ^{2}%
\left[ -dt^{2}+e^{2\lambda t}\left( dr^{2}+r^{2}d\Omega _{2}^{2}\right) %
\right] ,
\end{equation}%
by the following transformation%
\begin{eqnarray}
x_{1} &=&b_{0}\left( W_{0}-\left| W\right| \right) \left( \sinh \left(
\lambda t\right) +\frac{\lambda ^{2}r^{2}}{2}e^{\lambda t}\right) \\
x_{2} &=&b_{0}\left( W_{0}-\left| W\right| \right) \left( \cosh \left(
\lambda t\right) -\frac{\lambda ^{2}r^{2}}{2}e^{\lambda t}\right)  \notag \\
x_{3} &=&b_{0}\left( W_{0}-\left| W\right| \right) \lambda re^{\lambda
t}\cos \theta  \notag \\
x_{4} &=&b_{0}\left( W_{0}-\left| W\right| \right) \lambda re^{\lambda
t}\sin \theta \cos \varphi  \notag \\
x_{5} &=&b_{0}\left( W_{0}-\left| W\right| \right) \lambda re^{\lambda
t}\sin \theta \sin \varphi ,  \notag
\end{eqnarray}%
after some calculations we get the factor $b_{0}W_{0}\lambda $ in front of
the four dimensional part. This was already found as unity. Therefore the
four dimensional curved space time can be embedded in the five dimensional
flat space time by these coordinate transformations.

\subsubsection{$T_{55}\neq 0:$}

From the B-D equations we obtain that $p_{B}=-\rho _{B}\neq 0$ and $\beta
=0. $ As $\omega \rightarrow \infty ,$%
\begin{eqnarray}
\rho _{B} &=&-p_{B}\rightarrow \frac{\left( B_{0}c_{0}\right) ^{2/\alpha }}{%
2\left( b_{o}w_{o}\right) ^{2}}\frac{\alpha ^{2}\left( \alpha +1\right) }{%
\left( \alpha -1\right) }\phi ^{2-2/\alpha } \\
q_{B} &\rightarrow &\frac{\left( B_{0}c_{0}\right) ^{2/\alpha }}{\left(
b_{0}W_{o}\right) ^{2}}\frac{\alpha ^{2}}{\alpha -1}\phi ^{2-2/\alpha }
\notag \\
V_{0} &\rightarrow &\frac{\left( B_{0}c_{0}\right) ^{2/\alpha }}{\left(
b_{0}W_{0}\right) ^{2}}\frac{\alpha ^{2}\left( 3+\alpha \right) }{2\left(
\alpha -1\right) }.  \notag
\end{eqnarray}%
for all of the results $\beta =0$ and $V\left( \phi \right) =V_{0}\phi ^{2-%
\frac{2}{\alpha }},$ therefore the scalar potential becomes again time
independent for the exponentially expanding universe for $q_{B}\neq 0$.

If we suppose this phase as the false phase, the probability of a point
remaining in the false phase during the bubble nucleation process is quite
small as shown in \cite{Guth}. Then the universe is dominated by the true
vacuum and exits from the false vacuum.

In the true vacuum we can consider a power-law expansion.

\subsection{Power-law Expansion:}

The scale factors are:%
\begin{eqnarray}
a\left( t\right) &=&a_{0}\left( t/t_{0}\right) ^{\lambda }, \\
b\left( t\right) &=&b_{0}\left( t/t_{0}\right) ^{\gamma }.
\end{eqnarray}%
These power law solutions restrict us to choose $B\left( t\right)
=B_{0}\left( \frac{t}{t_{0}}\right) ^{\beta }.$ On the other hand B-D
equations is satisfied only if $\gamma ,\beta =1,$ and again we get same
result for the warp factor, \ $f\left( W\right) =\left( 1-\frac{\left|
W\right| }{W_{0}}\right) .$ Then these results causes scalar field to be
\begin{equation}
\phi \left( t,W\right) =B_{0}c_{0}\left[ \left( \frac{t}{t_{0}}\right)
\left( 1-\frac{\left| W\right| }{W_{0}}\right) \right] ^{\alpha },
\end{equation}%
where $B_{0}$ and $c_{0}$ are constants. Again to satisfy the B-D equations,
we find the similar scalar potential;%
\begin{equation}
V\left( \phi \right) =V_{0}\phi ^{\frac{2}{\alpha }\left( \alpha -1\right) },
\end{equation}%
and $B_{0}c_{0}$ has dimension $L^{-3/2}$, therefore $V_{0}$ has dimension $%
L^{-3/\alpha -2}$. Here to make B-D equations simpler we set $\frac{%
b_{0}W_{0}}{t_{0}}=1$.

Now we want to find a general result so we consider the equation of state as:%
\begin{equation}
p_{B}=\nu \rho _{B}.
\end{equation}%
Putting all of these settings in the B-D equations we find a nice result:
here the interesting thing is that there is no solution other than $\nu =%
\frac{1}{3}$ for $p_{B}\neq 0$ and $\rho _{B}\neq 0$ and solutions are valid
only for $q_{B}=0.$ Different values of the variables in equations (\ref{G00}%
-\ref{f}) are satisfied only for a single value of $\nu $ which is $\frac{1}{%
3}.$ Then this ratio between the pressure and energy density corresponds to
the radiation dominated universe; and $\omega $ dependence of $\lambda
,\alpha ,$ and $V_{0}$ are:

\begin{eqnarray}
\rho _{B} &=&3p_{B} \\
\alpha _{\pm } &=&\frac{\pm \sqrt{3\omega +4}+1}{2\left( \omega +1\right) },
\label{alfa} \\
\lambda _{\pm } &=&\frac{\omega \mp \sqrt{3\omega +4}}{4\left( \omega
+1\right) }
\end{eqnarray}%
and finally
\begin{equation}
V_{0\pm }=-\frac{3\left( B_{0}c_{0}\right) ^{2/\alpha }}{t_{0}^{2}}\frac{%
\left( 3\omega +4\right) \left( 3\omega \pm \sqrt{3\omega +4}+5\right) }{%
32\omega \left( \omega +1\right) ^{2}}.
\end{equation}

All of these solutions do not give a specific value for $\omega .$ From the
time-delay measurements, experimentally $\omega >500$ \cite{Reasenberg} and
more recently $\omega >3000$ \cite{Will}. As $\omega \rightarrow \infty ,$ $%
\alpha \rightarrow 0,$ $\lambda _{\pm }\rightarrow \frac{1}{4}$ and $%
V_{0}\rightarrow 0.$ This means that at this limit, the scalar field becomes
constant and the scalar potential vanishes.

For the power-law scale factor we obtain one more solution B-D equations
give the empty universe, namely $\rho _{B},$ $p_{B}=0$ and $\lambda
=1,V_{0}=0$ and $\alpha _{\pm }=\frac{\pm \sqrt{3\omega +4}+1}{2\left(
\omega +1\right) }$ which are \ the same as previous value of $\alpha $ (\ref%
{alfa}).

The solutions presented here represent decelerating cosmology for the
radiation dominated universe and expanding cosmology with constant velocity
for the empty universe. However astronomical observations show that the
universe is not only expanding but also undergoing accelerated expansion %
\cite{Riess,Page}. It may be possible to obtain power law acceleration in
B-D theory if scale factors for external dimensions are time dependent. In
string theory some cosmologies can achieve accelerating scale factors \cite%
{Townsend,Wohlfarth,Ohta}.

The metric which we found in this part may be related with \ the Kasner
space-time \cite{Kasner}. It has a cosmological singularity at $t=0$ where
the square of Riemann tensor diverges. On the brane we live ($W=0$)
\begin{equation}
R_{\mu \nu \rho \sigma }R^{\mu \nu \rho \sigma }=\frac{24\lambda ^{4}}{t^{4}}%
.
\end{equation}%
This is the physical singularity and it cannot be avoided by any coordinate
transformation \cite{Frolov}. However, since the central part of the
space-time is avoided in orbifold construction this has no importance for
the brane world scenario \cite{RS1}.

\section{The Effective Four Dimensional Gravitational Constant}

Finally we calculate the four-dimensional effective gravitational constant
on the brane and compare with the our previous results eq(\ref{g,mp,f}). On
the left hand side of the action in eq (\ref{action}) the first term is:%
\begin{equation}
\int d^{5}x\sqrt{g}\frac{\phi _{\left( 5\right) }^{2}}{8\omega }R_{\left(
5\right) }=\int d^{5}x\sqrt{g}M_{\left( 5\right) }^{3}R_{\left( 5\right)
}=\int d^{5}x\sqrt{g}\frac{1}{16\pi G_{\left( 5\right) }}R_{\left( 5\right)
}.
\end{equation}

We can perform the $W$ integral to obtain the effective gravitational
constant. With the same manner in the \cite{Bander} work, this equation
reduces to%
\begin{equation}
\int d^{5}x\sqrt{g}\frac{\phi _{\left( 5\right) }^{2}}{8\omega }R_{\left(
5\right) }=\int d^{4}xdW\sqrt{g^{\left( 4\right) }}\frac{\phi _{\left(
5\right) }^{2}}{8\omega }\left( 1-\frac{\left| W\right| }{W_{0}}\right)
^{2}b\left( t\right) R\left( g_{ij}^{\left( 4\right) }\left( x\right) \right)
\label{effective}
\end{equation}

For the exponentially increasing scalar factor we have obtained time
dependent scalar field that is $\phi _{\left( 5\right)
}=B_{0}c_{0}\left[
e^{\beta t}\left( 1-\frac{\left| W\right| }{W_{o}}\right) \right] ^{\alpha }$%
. Then eq(\ref{effective}) becomes%
\begin{equation}
\int d^{5}x\sqrt{g}\frac{\phi _{\left( 5\right) }^{2}}{8\omega }R_{\left(
5\right) }=\int d^{4}x2\frac{b_{0}W_{0}}{8\omega }\frac{%
B_{0}^{2}c_{0}^{2}e^{2\alpha \beta t}}{\left( 2\alpha +3\right) }\sqrt{%
g^{\left( 4\right) }}R\left( g_{ij}^{\left( 4\right) }\left( x\right) \right)
\end{equation}%
then since $\alpha \simeq 0,$ the effective gravitational constant becomes%
\begin{equation}
\frac{1}{16\pi G_{eff}}=M_{p\left( eff\right) }^{2}=\frac{\phi _{\left(
4\right) }^{2}}{8\omega }=\frac{b_{0}W_{0}}{12\omega }\left(
B_{0}c_{0}\right) ^{2},
\end{equation}%
which is independent of time. This is similar with the what we have
discussed in eq(\ref{g,mp,f}) and here $B_{o}c_{o}$ is within a few orders
of Planck mass.

For the power law scalar factors the scalar field is \bigskip $\phi _{\left(
5\right) }=B_{0}c_{0}\left( \frac{t}{t_{0}}\left( 1-\frac{\left| W\right| }{%
W_{0}}\right) \right) ^{\alpha }$%
\begin{equation}
\int d^{5}x\sqrt{g}\frac{\phi _{\left( 5\right) }^{2}}{8\omega }R_{\left(
5\right) }=\int d^{4}x\frac{b_{0}W_{0}}{4\omega }\left( \frac{t}{t_{0}}%
\right) ^{2\alpha +1}\frac{B_{0}^{2}c_{0}^{2}}{\left( 2\alpha +3\right) }%
\sqrt{g^{\left( 4\right) }}R\left( g_{ij}^{\left( 4\right) }\left( x\right)
\right) ,
\end{equation}%
here again for $\alpha \simeq 0,$ the four dimensional Brans-Dicke scalar
field (or the inverse of the effective gravitational constant) is%
\begin{equation}
\frac{1}{16\pi G_{eff}}=M_{p\left( eff\right) }^{2}=\frac{\phi _{\left(
4\right) }^{2}}{8\omega }=\frac{W_{0}}{12\omega }\left( B_{0}c_{0}\right)
^{2}b\left( t\right) .
\end{equation}%
Hence the four dimensional effective gravitational constant depends on time %
\cite{Bander}.

\section{Conclusion}

In this work we introduced a five dimensional B-D action and studied the
five dimensional metric with a warp factor. We showed that the field
equations imply a linear warp factor. For an inflating scale factor we found
that the energy density acts as an effective cosmological constant. For
power law expansion of scale factor we obtained a radiation dominated
universe. Additionally we have shown that the five dimensional scalar field
is nearly but cannot be exactly constant. On the other hand the four
dimensional effective scalar field is constant for exponentially growing
scale factor and depends on time for the power law scale factors.


\begin{thebibliography}{99}
\bibitem{Jordan} P. Jordan, Nature (London) 164, 637 (1949); Schwerkraft und
Weltall (Vieweg, Braunschweig, 1955); Z. Phys. 157, 112 (1959).

\bibitem{Brans} C. Brans and H. Dicke, Phys. Rev. \textbf{124}, 925 (1961).

\bibitem{Yasunory} Y. Fujii and K. Maeda, \textit{The Scalar--Tensor Theory
of Gravitation }(Cambridge University Press, 2003).

\bibitem{Banerjee} N. Banerjee and D. Pavon, Phys. Rev. D \textbf{63},
043504 (2001).

\bibitem{Sen} S. Sen and A.A. Sen, Phys. Rev. D \textbf{63}, 124006 (2001).

\bibitem{Arkani} N. Arkani-Hamed, S. Dimopoulos, and G.R. Divali, Phys.Lett.
B \textbf{429}, 263 (1998).

\bibitem{RS1} L. Randall and R. Sundrum, Phys. Rev. Lett. \textbf{83}, 3370
(1999).

\bibitem{RS2} L. Randall and R. Sundrum, Phys. Rev. Lett. \textbf{83}, 4690
(1999).

\bibitem{Freund} P.G.O. Freund, Nucl. Phys. B \textbf{209,} 146 (1982); L.
Qiang, Y. Ma, M. Han and D. Yu, Phys. Rev. D \textbf{71, }061501(R) (2005).

\bibitem{Mendes} L. E. Mendes ; A. Mazumdar, Phys. Lett. B \textbf{501,}
249-256 (2001).

\bibitem{potantial} J.M. Cline, H. Firouzjahi, Phys.Rev. \textbf{D64},
023505 (2001); S. Kobayashi and K. Koyama, J. High Energy Phys. JHEP \textbf{%
12}, 056 (2002); \ A. Mennim. R.A. Battye, Class.Quant.Grav. 18, 2171-2194
(2001); D. Langlois and M. Rodriguez-Martinez, Phys. Rev. D \textbf{64,}
123507 (2001); C. Barcelo and M. Visser, JHEP\textbf{\ 0010}, 019 (2000); C.
Barcelo and M. Visser, Phys. Rev. D. \textbf{63,} 0204004 (2000); K. Maeda
and D. Wands, Phys. Rev. D \textbf{62} 124009 (2000).

\bibitem{Bander} M. Bander, Phys. Rev. D \textbf{69}, 043505 (2004).

\bibitem{Deffayet} P. Binetruy, C. Deffayet, U. Ellanger and D. Langlois,
Phys. Lett. B \textbf{477} (2000) 269; P. Binetruy, C. Deffayet and D.
Langlois, Nucl. Phys. B \textbf{565} 269 (2000); S. M. Carroll, L. Mersini,
Phys.Rev. \textbf{D64} (2001) 124008; J. M. Cline, C. Grojean and G.
Servant, Phys. Rev. Lett. \textbf{83,} 4245 (1999); D. Langlois,
Prog.Theor.Phys.Suppl. \textbf{148}, 181-212 (2003); J.D. Arnowitt, B.
Dutta, Phys.Rev. \textbf{D70}, 126001 (2004).

\bibitem{Weinberg} E. J. Weinberg, Phys. Rev. D \textbf{40}, 3950 (1989).

\bibitem{Daila} D. La and P.J. Steinhardt, Phys. Rev. Lett. \textbf{62}, 376
(1989).

\bibitem{biz} M. Ar\i k \textit{et al.}, Phys. Rev. D \textbf{68}, 123503
(2003).

\bibitem{Guth} A. H. Guth, Phys. Rev. D \textbf{23}, 347 (1981).

\bibitem{Reasenberg} R.D. Reasenberg \textit{et al.}, Astrophys. J. \textbf{%
234}, L219 (1979).

\bibitem{Will} C. M. Will, \textit{Theory and Experiment in Gravitational
Physics (}Cambridge University Press, 1993), revised ed., p. 123; in \textit{%
Proceedings of the 1998 Slac Summer Institute on Particle Physics }%
(gr-gc/9811036).\textit{\ }

\bibitem{Townsend} P.K. Townsend and M.N.R. Wohlfarth, hep-th/0303097.

\bibitem{Wohlfarth} M.N. Wohlfarth, hep-th/03071719.

\bibitem{Ohta} N. Ohta, Phys. Rev. Lett. \textbf{91, }061303 (2003).

\bibitem{Riess} A.G. Riess \textit{et. al., }Astrophys. J. \textbf{560, }49
(2001).

\bibitem{Page} L. Page, astro-ph/0306381.

\bibitem{Kasner} E. Kasner, Am. J. Math. \textbf{43, }126 (1921); \textbf{43,%
} 217 (1921).

\bibitem{Frolov} A. V. Frolov, Phys. Lett. B \textbf{514}, 213 (2001).
\end{thebibliography}
\end{document}